# Research on spatial information transmission efficiency and capability of safe evacuation signs

Ruiwen Fan [a], Zhangyin Dai [a], Shixiang Tian [a], Ting Xia [a], Hui Zhou [a], Congbao Huang [a]

[a] Mining College, Guizhou University, Guizhou, Guiyang 550025 China

RuiwenFan, E-mail addresses:starfrw@163.com
Corresponding author.
E-mail addresses: zydai1@gzu.edu.cn(Zhangyin Dai)

**Highlights**

- Experiments are carried out by a combination of simulation experiments and fire drills.
- Different genders are affected differently by changes in the angle of safety evacuation signs.
- A formula for calculating the visible area of safety evacuation signs is proposed.
- Design space three-dimensional structure safety evacuation signs.

**Abstract:** As an indispensable spatial direction information indicator for emergency evacuation, the spatial relationship between safety evacuation signs and evacuees will affect the response time of evacuees and the evacuation efficiency. This paper takes 2 kinds of common safety evacuation signs, hangtag-type and embedded, as the research object and designs space direction information transmission efficiency and capability simulation experiment and fire drill, the efficiency and capability of spatial direction information transmission of safety evacuation signs are studied. The results show that the space angle of the hangtag-type safety evacuation sign is inversely proportional to the information transmission efficiency and capability of the space direction, and the fire drill also confirms this conclusion. When the spatial angle of the embedded safety evacuation sign is 5°, the spatial direction information transmission efficiency and capability increase. Simultaneously, the average escape time of the participants in the fire drill was lower, and the percentage of choosing unfamiliarity exports increased. The evolution of spatial angle has no significant effect on the intention of the response of subjects of different genders; when choosing the direction, males are more easily affected by the change of spatial angle than females; the confidence level of females' choice is more easily affected by spatial angle. In addition, according to the research results, the corresponding three-dimensional structure safety evacuation signs are designed. The functional structure of the safety evacuation signs is perfected, which can effectively improve the efficiency of fire emergency evacuation.

## 1 Introduction

At the beginning of the 21st century, among the 63 million people globally, the annual death toll was about 70,000 to 80,000, and the number of fire injuries was about 5,000 to 8,000[1]. When a fire occurs, it is difficult for evacuees to find a suitable evacuation path due to psychological pressure, time constraints, and visibility caused by fire[2][3]. Therefore, additional spatial information is needed to help them find safe exits[4]. As an essential part of the emergency evacuation system and the primary medium for transmitting spatial direction information, the safety evacuation sign impacts evacuees' accurate and rapid escape in a fire. However, its actual effect is not ideal, proven in previous disasters[5][6] and experimental studies[7][8]. Moreover, when the subject is in different positions of the safe passage, the spatial angle between the line of sight and the safe evacuation sign will change. Sometimes it is difficult to obtain the spatial direction information conveyed by the safety evacuation signs in time, which increases the evacuation time. Therefore, an in-depth study of the information transmission efficiency and capability of the safety evacuation signs space direction is of great significance to improve the efficiency of an emergency evacuation.

The information transmission efficiency and capability of space direction of safe evacuation signs has always been a research hotspot and difficulty in the field of an emergency evacuation. Many scholars have improved the intelligence of fire safety evacuation systems. For example, a new type of intelligent safety evacuation sign[9][11] or a variable forbidden exit sign[12] can be used to plan the optimal route according to the real-time fire situation to achieve the goal. Existing studies have shown that new intelligent safety evacuation signs positively impact evacuation efficiency[13]-[14]. Although the research on new intelligent safety evacuation signs has gradually matured, traditional safety evacuation signs are currently the most widely used. The research on the information transmission efficiency of traditional safety evacuation signs in spatial direction is mainly carried out from two aspects: information perception[15]-[17] and appearance structure.

In terms of information perception, research-based on human vision, hearing, touch, etc., have gradually developed[18]-[21]. Among them, how to improve the visual appeal of safety evacuation signs has always been the research focus of information perception. Scholars mostly start from the flickering state of safety evacuation signs[22], color, size, etc.[23][24]. As demonstrated by Nilsson et al.[25][26], adding a flashing green light above the emergency exit makes it more likely that the emergency exit will be used in evacuation trials. Collins et al.[27] found that the visibility of emergency evacuation signs is related to uniformity, brightness, and contrast. Some scholars have also conducted experiments on the four most commonly used visual variables of size, hue, brightness, and direction, and found that size is the most effective visual variable, and direction is the most ineffective visual variable in flickering patterns[28][29]. Jeon et al.[30] started by changing the material of the safety evacuation signs and found that when the installation interval of the phosphorescent signs is smaller than the human stride, it is helpful to improve the safety evacuation efficiency. Nassar et al.[31] used a simulation model to provide a framework for rational positioning of signs in public spaces by considering the geometry of the space, the travel patterns of passengers or pedestrians, the location of obstacles, the legibility distance, and sign design to improve its visibility. Occhialini et al.[32] quantitatively assessed the perception ability of safety evacuation signs in virtual environments based on neural activity analysis, especially the effectiveness of safety evacuation sign types and locations. Other scholars have studied the effects of distance between safety evacuation signs and evacuees[33]-[35], space configuration[34], visibility[36][38], etc., on the visual appeal of safety evacuation signs. With the in-depth development of research, Xie[39], Kubota[40], and others introduced the concept of interaction angle, showing that it can affect the direction choice of evacuees, providing a new idea for the study of visual attractiveness of safe evacuation signs.

The study of appearance structure is also more detailed. In addition to the research on optimizing the appearance size of the common cuboid safety evacuation signs[41-42], some scholars have studied and designed other shapes of safety evacuation signs to improve the efficiency of spatial direction information transmission efficiency. For example, Liu et al.[43] developed a polyhedral evacuation channel intersection induction sign. Wu[44] and Mou[45] designed a three-dimensional space embedded safety evacuation sign in the shape of an isosceles triangle in utility model patents CN209880078 and CN212516433.

The above research results can effectively reduce the evacuation time and improve evacuation efficiency. However, the spatial structure of safety evacuation signs, especially the influence of space angle on the information transmission efficiency and capability of safety evacuation signs, is rarely considered. This work aims to study the effect of the spatial angle of safety evacuation signs on spatial orientation information transmission effectiveness. Firstly, the simulation experiment of spatial direction information transfer efficiency and capability is carried out. This is to analyze the influence of the spatial angle of the two types of common safety evacuation signs in the building on the respondent's choice of response, the expected choice conformity, and the choice confidence level. Afterwards, fire drills are carried out to verify the accuracy of the simulation experiment of spatial direction information transfer efficiency and capability. In addition, according to the research results, the safety evacuation signs of the three-dimensional space structure are designed to provide new ideas for the research on the visual attractiveness of safety evacuation signs.

## 2  Experiment brief

In this paper, the PsyLab psychological experiment system is used to carry out the choice response time experiment, and the purpose is to evaluate the testee's response speed. Then use the response time timing software designed based on the PsyLab psychological experiment system to conduct a simulation experiment of spatial direction information transfer efficiency and capability (hereinafter referred to as Experiment 1). The purpose is to analyze the influence of the safety evacuation signs at different spatial angles on the respondent, the degree of conformity of the expected choice (the direction indicated when the spatial angle α is 0° is the expected direction) degree of confidence in the choice. The ability of safe evacuation signs in different spatial angles to transmit spatial direction information is obtained. Then a fire drill (hereinafter referred to as Experiment 2) was carried out to verify whether the conclusions of Experiment 1 were realistic. Among them, the second experiment was carried out in the natural environment of a comprehensive office building of a university. The real-world scene is reproduced in 3D renderings and used for experiment 1. The purpose is to ensure the consistency of Experiment 1 and Experiment 2's consistency and to reduce errors caused by external factors.

### 2.1  Selecting a response time experiment

Response time is the time interval between receiving an external stimulus and responding. Response time can be divided into 3 types; simple response time, identification response time, and selection response time [46]. In 1865, Donders was the first scientist to measure response times in the laboratory, noting that simple response times were faster than identification response times, and selection response times were longer than simple and identification response times[47]. To a certain extent, the response time reflects the psychological characteristics such as people's adaptability and attention characteristics and can show the individual's cognitive ability.

In Donders' choice-response experiment, different stimuli are presented at random. Subjects need to identify which stimulus is present and respond appropriately[47]. This paper selects the red and green circles to stimulate the subjects. The purpose of the choice-response experiment is to evaluate whether the response speed of the subjects deviates from the group and select the subjects with minor differences in response speed to participate in the following experiment to reduce the error caused by the difference in the subjects' response speed.

Sixty-three subjects were recruited in this college, including 45 males and 18 females, which conformed to the gender ratio of engineering disciplines in China[48]. They are all young students of Chinese nationality, ranging in age from 18 to 23 years old (the average age is 20 years). The choice response time experiment is based on the PsyLab psychological experiment system. The system contains more than 100 psychological experimental principles and methods, such as response time experiments, perception experiments, etc. Moreover, experiments can be carried out with the help of a computer platform. During the experiment, the subjects observed the stimuli of the red and green circles presented by the computer and pressed the corresponding buttons, and the system automatically counted when the subjects chose to respond. The experimental process is shown in Fig.1.

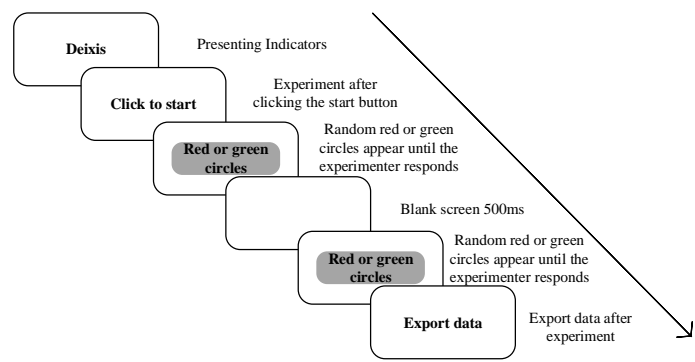

Fig.1  Experimental flow chart for selecting reactions

### 2.2  Experiment 1
#### 2.2.1  Experiment platform

Based on the PsyLab psychological experiment system, C# is used for independent programming, and a response time timing software that can automatically record and save the response time of the testee is designed and developed. In order to test the subject's choice response after seeing the stimulus picture, the use process is shown in Fig.2. The subjects who participated in the experiment were the same as those who chose the responses. Experiment 1 were conducted in quiet classrooms. The experimental stimuli are presented through the designed experimental platform. The monitor screen size is 22 inches, and the resolution is 1920×1080. During the experiment, the subject's position is fixed, and the distance from the screen is about 600mm.

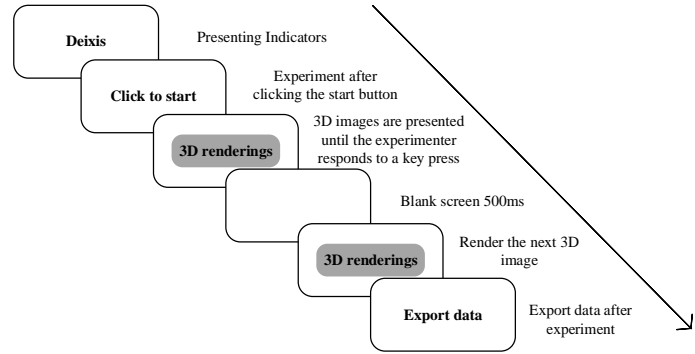

**Fig.2  Flowchart for the use of timing software in response time**

According to China's "*Fire Safety Signs*"[49], "*Fire Safety Signs Setting Requirements*"[50], and other relevant regulations, the size of the safety evacuation signs in the 3D renderings is 359×149×23mm. This is the same size as China's common safety evacuation signs (as shown in Fig.3). The height of the sign from the ground meets the relevant regulations. Referring to the research of Kubotaet al.[40], considering that when $α_d=90°$, the subject can only observe the side of the safety evacuation sign, so the space angle αd of the hangtag-type safety evacuation sign is selected as 0°, 30°, 60°, 70°, 80°, as shown in Fig.4. Considering the protruding wall height and other factors, the space angle $α_q$ of the embedded safety evacuation sign is determined to be 0°, 3°, 5°, 10° ($α_q=0°$ when the embedded safety evacuation sign is parallel to the wall), as shown in Fig.5.

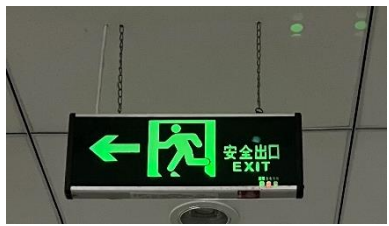 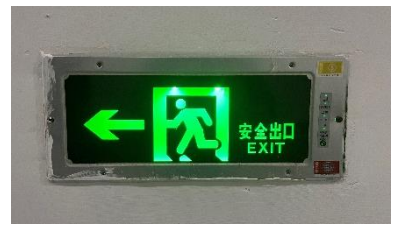

(a) Hangtag-type safety evacuation sign　　　　　(b) Embedded safety evacuation sign

**Fig.3  Safety evacuation sign**

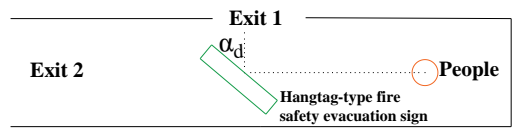

**Fig.4  Schematic diagram of the spatial angle of hangtag-type safety evacuation sign**

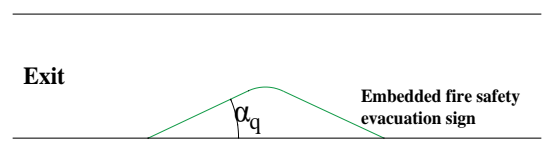

**Fig.5  Schematic diagram of the spatial angle of embedded safety evacuation sign**

The 3D rendering scene and background color are consistent with the more familiar university office building. Among them, three types of familiar safety passages are selected for the evacuation scene of the hangtag-type safety evacuation sign, including Z-type, L-type, and F-type (as shown in Fig.6). Embedded safety evacuation signs choose familiar linear safety passages for evacuation scenarios. The subject's eye height is 1500mm according to "*Chinese Adult Body Dimensions*" (GB/T 10000-1988), and the distance from the safety evacuation sign is 2000mm. Using 3DMAX software, a total of 19 3D renderings of the evacuation scene were drawn, of which the 3D renderings of the evacuation scene of the hangtag-type safety evacuation sign included 5 different spatial angles (0°, 30°,60°, 70°, 80°) under the Z-type, L-type, and F-type safety passages, a total of 15. The 3D renderings of the emergency evacuation scene with embedded safety evacuation signs include four different spatial angles (0°, 3°, 5°, 10°).

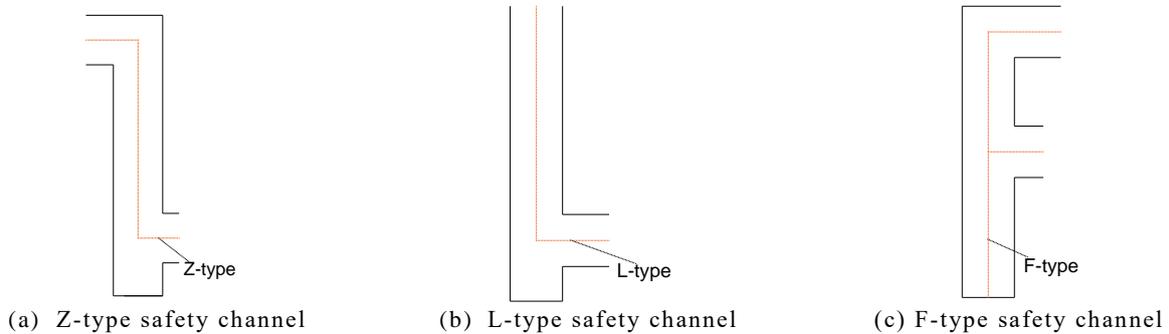

(a)　Z-type safety channel　　　　(b)　L-type safety channel　　　　(c)　F-type safety channel

**Fig.6　Schematic diagram of hangtag-type safety evacuation sign evacuation scene**

Fig.7 shows the 3D renderings of Z-type, L-type, and F-type safety passages from different spatial angles in the evacuation scene of the hangtag-type safety evacuation sign. Considering the length of the article, the 3D renderings from other spatial angles will not be displayed one by one.

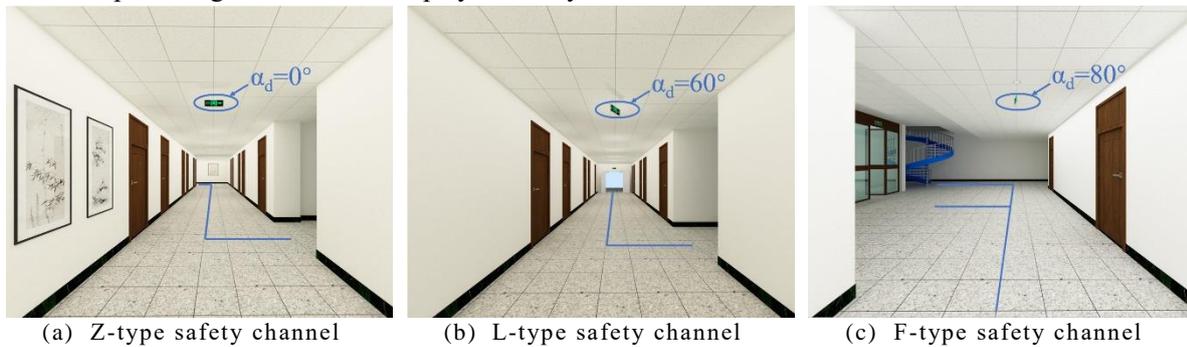

(a)　Z-type safety channel　　　　(b)　L-type safety channel　　　　(c)　F-type safety channel

**Fig.7　3D rendering of evacuation scene with hangtag-type safety evacuation sign**

The 3D rendering of the embedded safety sign's evacuation scene is shown in Fig.8.

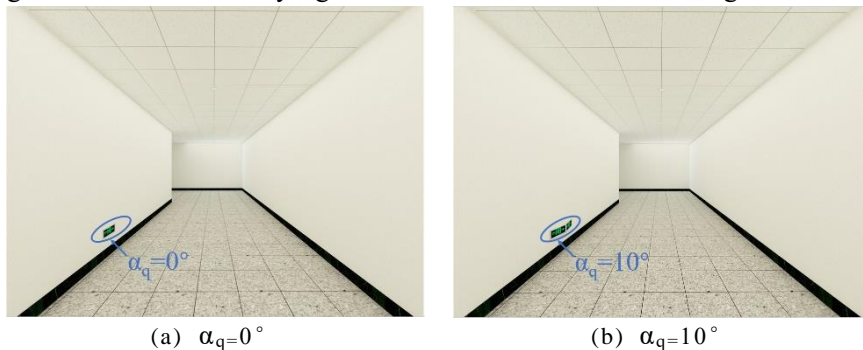

(a)　$\alpha_{q}=0°$　　　　(b)　$\alpha_{q}=10°$

**Fig.8　3D rendering of evacuation scene with embedded safety evacuation sign**

### 2.2.2　Experimental procedure

Experiment 1 was carried out based on the response time timing software, and the specific steps were as follows:

1) Explain the experimental content and operation instructions to the subjects to understand the experimental content, process, and requirements. The experiment aims to study "people's responses when they see safety evacuation signs from different spatial angles".

2) The test subject sits in front of the computer, imaginary in an emergency fire evacuation scene. Click to enter the response time timing software, double-click the blank space after reading the instructions, and 19 3D renderings of different scenes and spatial angles will be displayed on the computer screen. There is a blank screen for 500ms between every two 3D renderings, and the subject must choose the safe exit direction as soon as possible. The system is equipped with two buttons, and the subjects have been informed of the meaning of each button before the experiment.

3) After completing the direction selection, the subjects were asked to use a 4-point scale (1=not at all confident, 2=not sure, 3=confident, 4=very confident) to dictate the degree of confidence in the choice of the safe exit direction for the scene, and the staff members record.

4) At the end of the experiment, export the test data.

### 2.3 Experiment 2
#### 2.3.1 The fire drill site

The second place of the experiment is a comprehensive office building of a university, with five floors. The office building has one elevator, but the participants were informed in advance that it was forbidden to use it. There are two stairs on each floor for escape. There are four safety exits, namely the main entrance on the first floor, the two side doors, and the glass door on the second floor, which are named No.1, No.2, No.3, and No.4 safety exits shown in Fig.9. The No.1 safety exit is a typical safety exit for the participants, and the other safety exits are usually closed and will only be opened in case of fire and other emergencies. Before the experiment, remind the participants that all safety exits will be opened for escape.

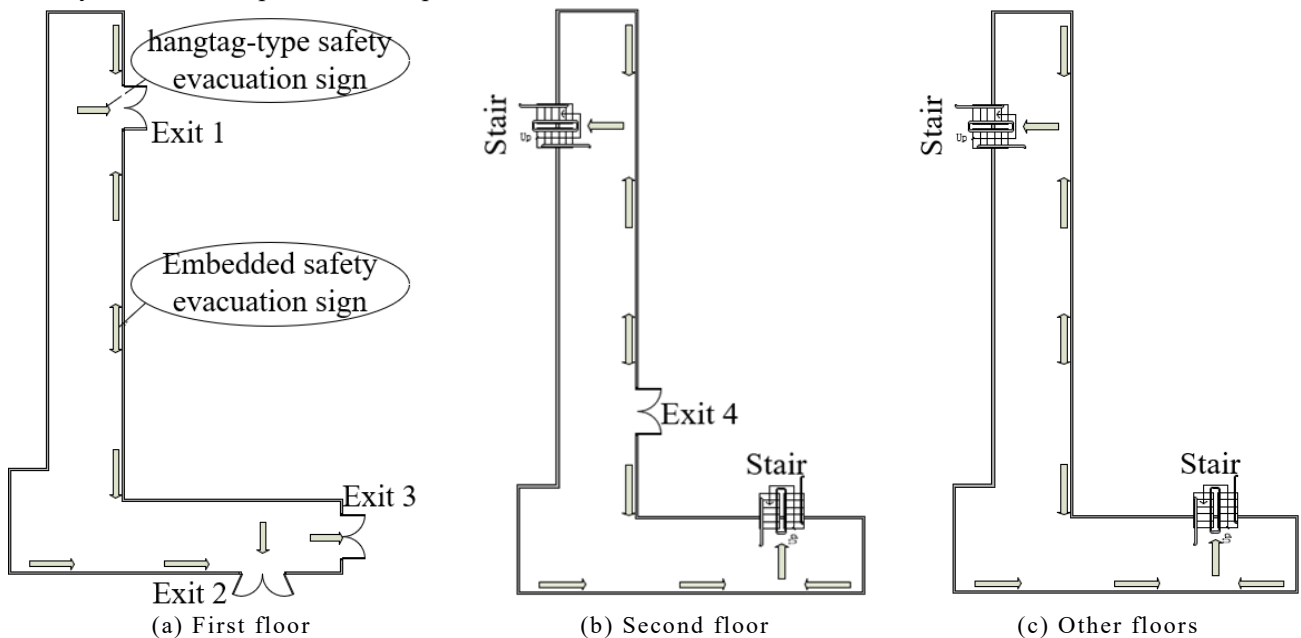

Fig.9 Schematic diagram of floor safe passage

#### 2.3.2 Experimental procedure

The participants in the second experiment were the same as those in the previous experiment (individual personnel was absent). Since they are all students of this college, they have all been in and out of the college building. Through the interview before the experiment, it was found that the participants had all evacuation experience before. In order to make the participants more familiar with the exercise environment and procedures, before the official start of the exercise, organize all participants to conduct a rehearsal. The specific steps of the exercise are as follows:

1) Explain the exercise's content, procedures, and requirements to the participants to ensure that they understand that the purpose of the exercise is to study the impact of embedded safety evacuation signs from different spatial angles on the escape speed of personnel.

2) Personnel were randomly assigned to five groups, one corresponding to one floor. Before the exercise, the participants stood scattered on the corresponding floors, as shown in Fig.10.

3) When the fire alarm sounded, the participants began to escape in an escape posture (as shown in Fig.11), and the staff at each exit used a stopwatch to time the time. The stopwatch used has passed the ISO quality management system certification, and the timing unit is 1/100s.

4) After each fire drill, the participants rest for 15 minutes, and the staff changes the space angle of the safety

evacuation sign. After the completion, the participants will change the floor, and the next fire drill will start, and the participants will be evacuated until the fire drill is over.

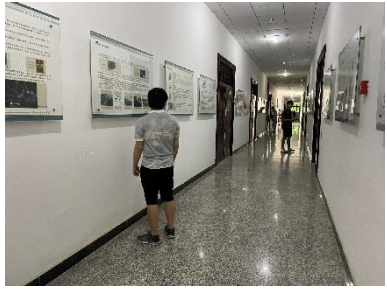 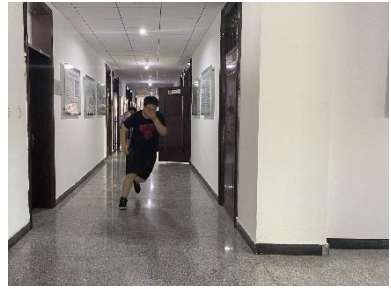

Fig.10  Floor distribution map of drill personnel　　Fig.11  Drill personnel escape scene diagram

### 2.4  Data processing method

In this paper, Tukey's test method[51] is used to process outliers for all the original data of Experiment 1 and Experiment 2. The method is to use quartiles to identify outliers. First, arrange the sequence from small to large, and remove the quartile $Q_1$ and the upper quartile $Q_3$. Then, those less than $Q_1$-k $(Q_3-Q_1)$ or greater than $Q_3$+k $(Q_3-Q_1)$ are outliers. Where k is a parameter. Generally speaking, k is 1.5 to be the range of moderate outliers, and 3 is the range of severe outliers. In this paper, the value of k is 1.5, and the median replaces the detected outliers. The Shapiro-Wilk test was performed on experiment 1 and the data of experiment 2 after outlier treatment. After checking the normality of the distribution, it was found that they were all non-normal distributions. Therefore, based on SPSS 22.0, Friedman's M test was used, and the difference between multiple paired samples was carried out with p=0.05 as the significance level. The processed data are analyzed by statistical methods, presented in graphs, and explained by text.

## 3  Results and Analysis

### 3.1  Selecting a response time experiment

The number of samples in this experiment is small, and random errors in the data are unavoidable. In order to reduce the error caused by outliers, this paper adopts the principle of Tukey's test to process outliers in the data. Outliers smaller than $Q_1$-1.5IQR or larger than $Q_3$+1.5IQR were removed ($Q_1$ is the lower quartile, $Q_3$ is the upper quartile). The box plot of the data when the reaction is selected is shown in Fig. 12.

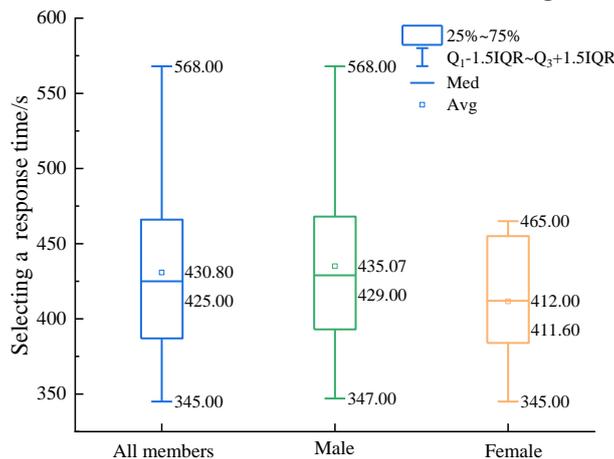

Fig.12  Box plot of select **response** time

It can be seen from Fig.12 that there is no maximum or minimum value deviating from the overall data in the experimental data. It shows that there is little difference in response speed between each subject, and all subjects can enter the next stage of the test. And when analyzed by gender, the response time eigenvalues of women are all smaller than those of men, that is, in the simple choice response time test, women's response time is faster.

### 3.2  Hangtag-type safety evacuation sign

#### 3.2.1　　Experiment 1

1)Z-type safety channel

Fig.13 shows the effect of spatial angle on the subjects' choice of response.

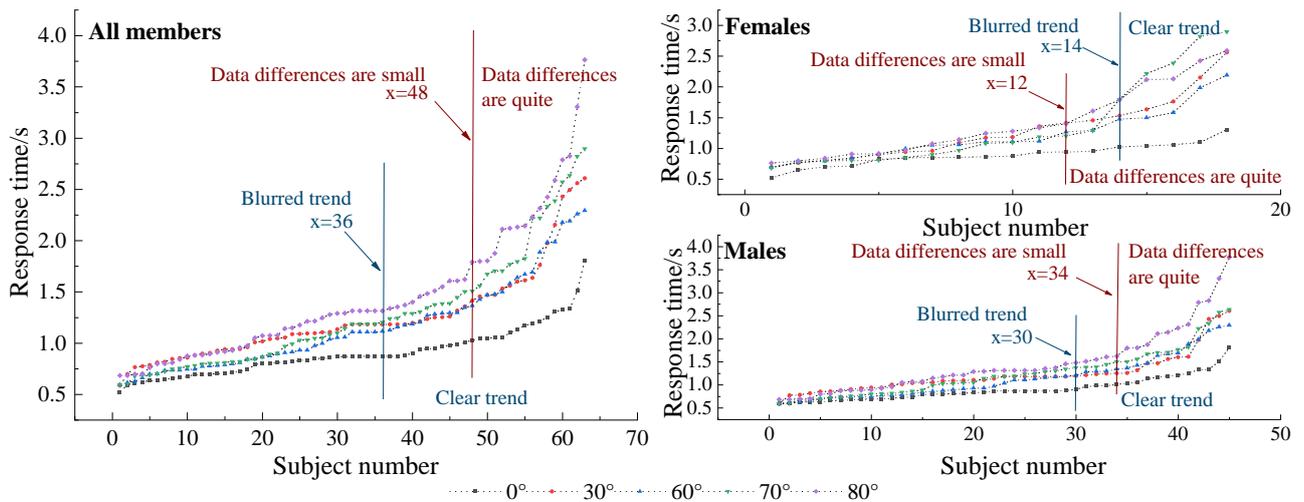

Fig.13  Selecting the response time to change with space angle contrast

In the Z-type scene, no statistically significant difference was found between 30° and 70° (p=0.91), while the results for all other spatial angle pairings were highly statistically significant (p=0). As shown in Fig.13, when all members were analyzed, the subjects' response time was the shortest when the spatial angle was 0°, and the longest when the spatial angle was 80°. When the number of subjects tested exceeds 36, this trend is more apparent. Among them, the data of 15 people's choice response times were quite different from others. When analyzed by gender, the trend is consistent with the overall analysis. Among them, the number of male subjects is more than 30, the trend is more obvious, and the number of female subjects is more than 14. 11 male subjects and 6 female subjects differed from the others when they responded. That is to say, the overall response time of the subjects is positively correlated with the spatial angle. When the spatial angle is 80°, the subject's response time is the largest. From a spatial perspective, there is little difference between males and females on the influence of the subjects on the choice of responses.

Fig.14 shows the effect of spatial angle on the subject's choice of direction.

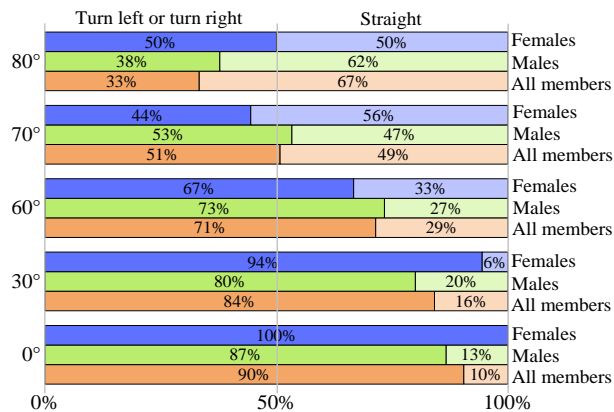

Fig.14  Direction selection and space angle quantitative relationship

As shown in Fig.6, the first exit in the Z- type safe passage has 2 options: right (expected direction) and straight (unexpected direction). As shown in Fig.14, analyzing the entire membership, fewer and fewer people choose the expected direction as the spatial angle increases. When analyzing gender, the trend of male subjects' selection was consistent with the overall trend. The proportion of female subjects choosing the expected direction is slightly higher than that of male subjects when the spatial angle is 0°, 30°, and 80°. When the spatial angle is 60° and 70°, it is slightly lower than that of males. That is to say, the proportion of subjects choosing the expected direction as a whole is negatively correlated with the spatial angle, and the proportion of women choosing the expected direction is higher than that of men as a whole. From the analysis of the influence of the spatial angle on the direction choice of the subjects, men are more likely to be affected by the change of the spatial angle than women.

Fig.15 shows the effect of spatial angle on the subjects' choice confidence.

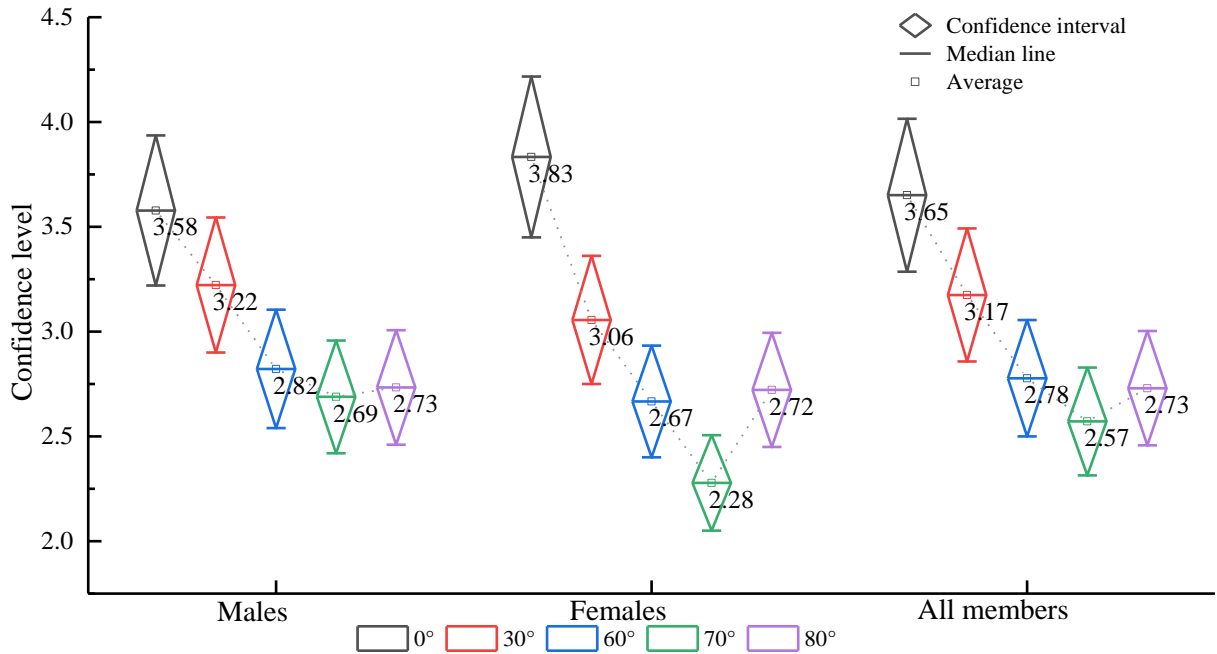

**Fig.15 Degree of confidence and space angle quantitative relationship**

This experiment uses a 4-point scale to evaluate the subjects' understanding of the spatial orientation information provided by the safety evacuation signs. As can be seen from Fig.15, when the spatial angle is 0° and 30°, the average self-confidence degree of the subjects is more than 3; when the spatial angle is more than 60°, the average self-confidence degree is less than 3, and the self-confidence degree when the spatial angle is 80° will rise slightly. However, the confidence level of female subjects is much higher than that of men when the spatial angle is 0°, and the confidence level of female subjects is much lower than that of men when the spatial angle is 70°. When women change their spatial perspective, their choice confidence level will decrease.

2)L-type safety channel

Fig.16 shows the effect of spatial angle on the subjects' choice of response.

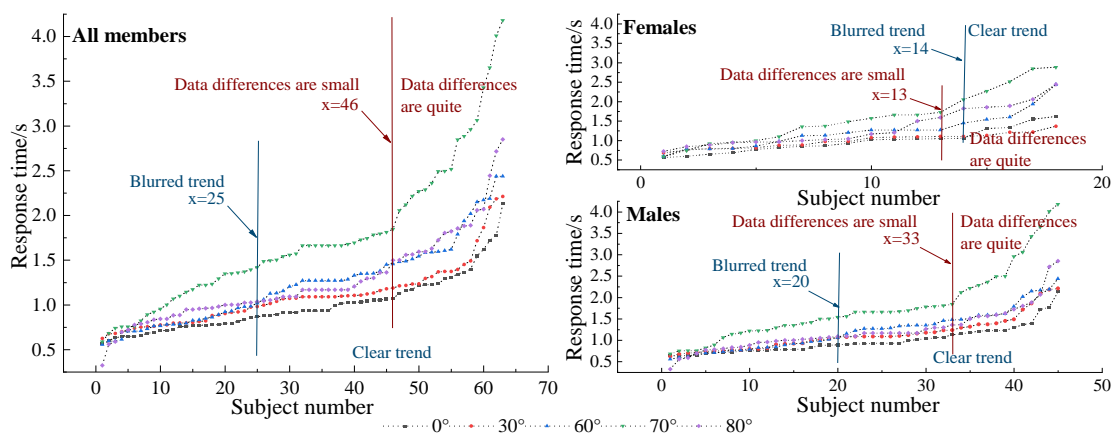

**Fig.16 Selecting the response time to change with space angle contrast**

In the L-type scene, no statistically significant difference was found between 60° and 80° ($p=1.00$), and the results of other spatial angle pairings were all statistically significantly different ($p=0$), where, When 30°vs60°, $p=0.016$. As shown in Fig.16, when all members are analyzed, when the spatial angle is 0°, the respondent's response time is the shortest; when the spatial angle is 70°, the respondent's response time is the longest, and when the spatial angle is 80°; When the number of subjects exceeds 25, this trend is more clear; and the data of 17 of them are pretty different from those of others. When analyzed by gender, the changing trend is consistent with the overall analysis; the trend is more obvious when the number of male subjects is more than 20 and the number of female subjects is more than 14; there are 12 male subjects and 5 female subjects. The tester's response time is quite different from that of others. That is to say, on the whole, there is a positive correlation between the response time of the subjects and the spatial angle, but when the spatial angle is 70°, the response time of the subjects is the largest. From a spatial perspective, there is little difference between males and females in the influence of the subjects on the choice of responses.

Fig.17 shows the effect of spatial angle on the subject's choice of direction.

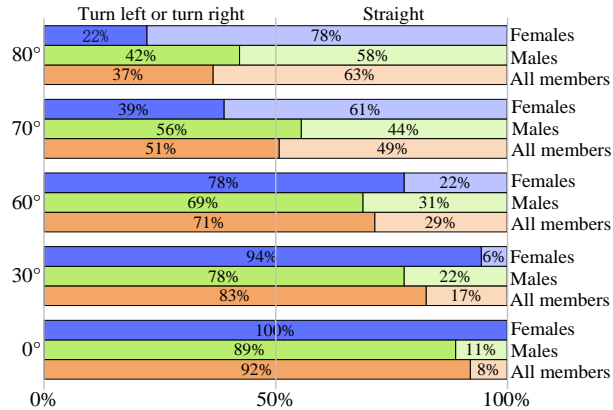

**Fig.17 Direction selection and space angle quantitative relationship**

Consistent with the Z-Type safety channel, the subject at the first exit in the L- type safety channel also has 2 choices: right (expected direction) and straight (unexpected direction). As shown in Fig.17, the analysis of all members shows that as the spatial angle increases, fewer and fewer people choose the expected direction. When analyzed by gender, male respondents' selection trend is consistent with the overall trend. The proportion of female respondents choosing the expected direction is slightly higher than that of male respondents when the spatial angle is 0°, 30°, and 60°. The spatial angle is at 70° and 80°, slightly lower than males. That is to say, the proportion of subjects choosing the expected direction as a whole is negatively correlated with the spatial angle, and the proportion of women choosing the expected direction is higher than that of men as a whole. Men are more susceptible to changes in spatial angle than women.

Fig.18 shows the effect of spatial angle on the subjects' choice confidence.

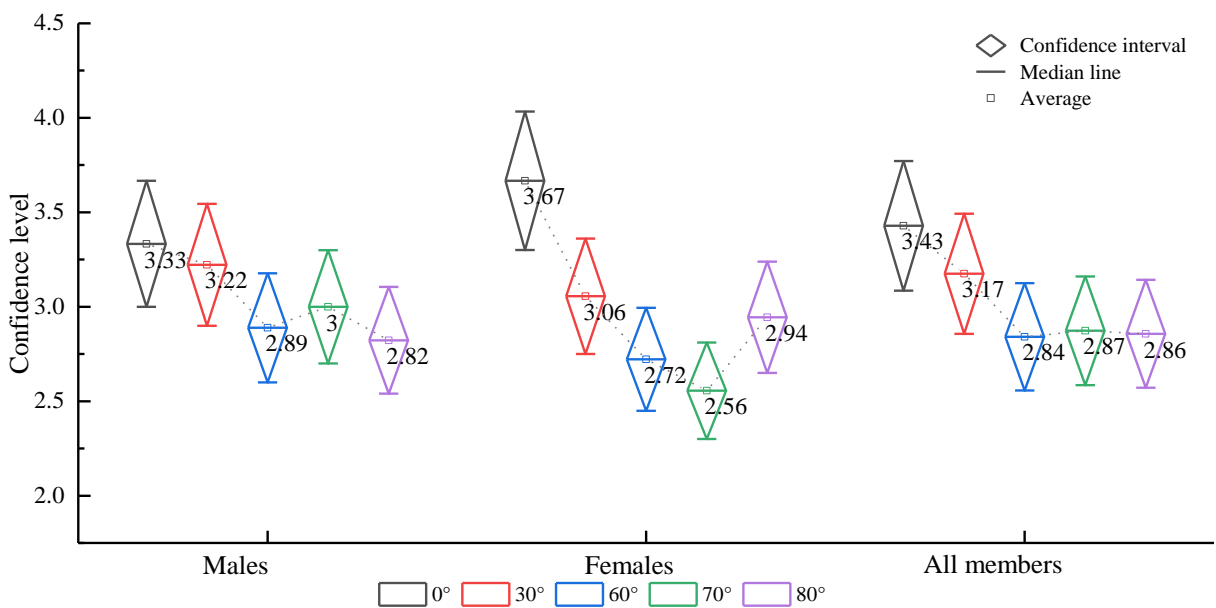

**Fig.18 Degree of confidence and space angle quantitative relationship**

It can be seen from Fig.18 that in the L-type safety channel as a whole, when the spatial angle is 0° and 30°, the average self-confidence degree of the subjects is greater than 3. When the spatial angle is greater than 60°, the average confidence level is less than 3. However, the confidence level of male subjects when the spatial angle is 70° is slightly higher than that when the spatial angle is 60° and 80°, and the confidence level of female subjects when the spatial angle is 80° is slightly higher than that when the spatial angle is 80° at 60° and 70°. As the spatial angle increases, the overall level of confidence in the test taker's choice decreases, and the level of confidence in the test taker's choice is the same when the spatial angle is 60°, 70°, and 80°. The final result is that women's choice confidence level will decrease when the spatial angle changes.

3)F-type safety channel

Fig.19 shows the effect of spatial angle on the subjects' choice of response.

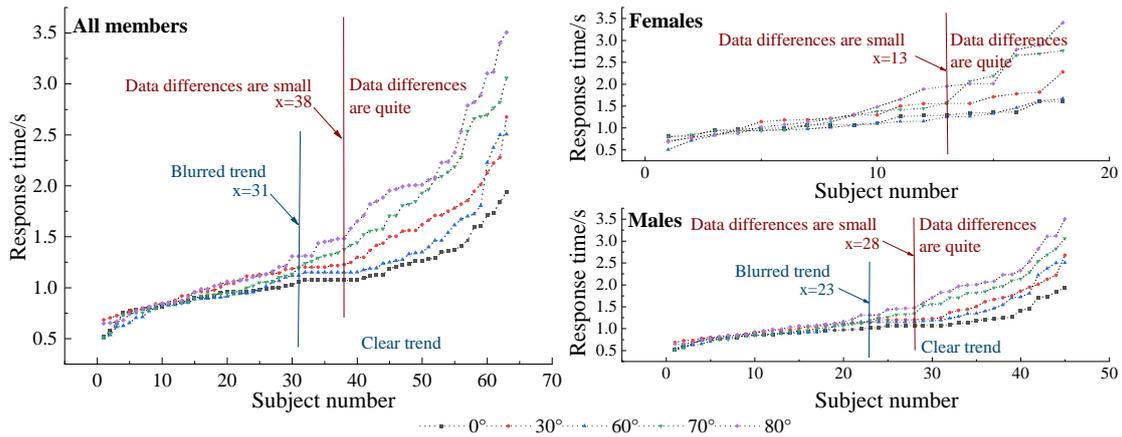

Fig.19 Selecting the response time to change with space angle contrast

No statistically significant differences were found between the 0° vs 60°, 30° vs 70° conditions (p=1) in the F-type scenario, while all other spatial angle pairings were highly statistically significantly different (p=0), where, at 30°vs80°, p=0.004. As shown in Fig.19, when all members are analyzed, when the spatial angle is 0°, the subjects choose the shortest response time. When the spatial angle is 80°, the subject responds the longest. When the number of subjects tested exceeds 31, this trend is clearer. Among them, 25 people's choice response data are quite different from others. When analyzed by gender, the trend is consistent with the overall analysis. The trend is more obvious when the number of male subjects is more than 23, while the overall trend of female subjects is more ambiguous. There were 17 male subjects and 5 female subjects who were different from the others when they responded. There is a positive correlation between the subject's choice of response and the spatial angle. From the perspective of the influence of the spatial perspective on the respondents' choice and response, there is little difference between males and females, but female respondents are more sensitive to changes in spatial perspectives.

Fig.20 shows the effect of spatial angle on the subject's choice of direction.

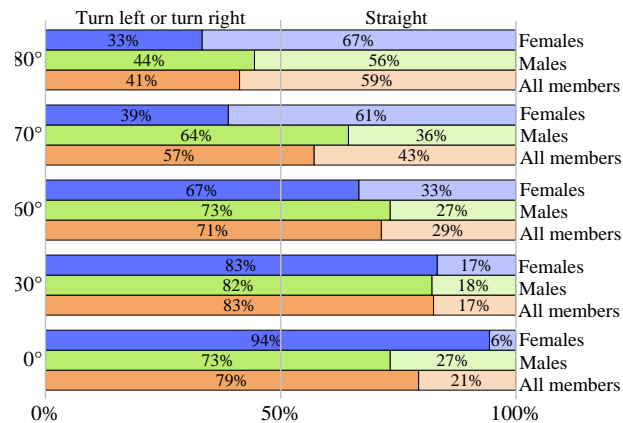

Fig.20 Direction selection and space angle quantitative relationship

Unlike the Z-type and L-type safety channels, the 2 safety exits in the F-type safety channel are located on the left side of the subject's line of sight and are close to each other. Therefore, the subjects were reminded before the experiment that the near safe exit is left (expected direction), and the far safe exit is straight (unexpected direction). As shown in Fig.20, the analysis of all members shows that as the spatial angle increases, the proportion of the subjects who select the expected direction gets lower and lower. When analyzed by gender, the selection trends of male and female subjects were consistent with the overall, and the proportion of female subjects who chose the expected direction was slightly higher than that of male subjects when the spatial angle was 0° and 30°, and slightly lower than that of males when the spatial angle was 60°, 70°, and 80°. That is to say, the proportion of the subjects who select the expected direction is negatively correlated with the spatial angle. From the analysis of the influence of the spatial angle on the direction choice of the subjects, the proportion of women choosing the expected direction is generally higher than that of men.

Fig.21 shows the effect of spatial angle on the subjects' choice confidence.

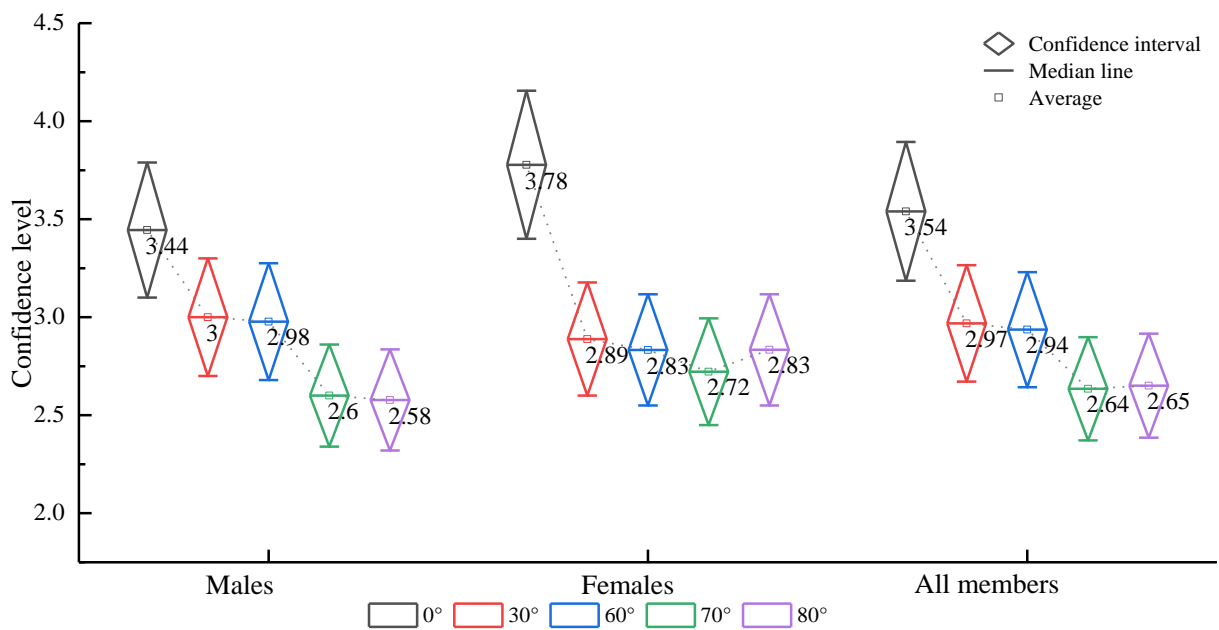

Fig.21 Degree of confidence and space angle quantitative relationship

It can be seen from Fig.21 that, consistent with the Z-type and L-type safety channels, in the F-type safety channel, with the increase of the spatial angle, the subjects' confidence in their own choices shows a decreasing trend. In the F-type safety channel as a whole, when the spatial angle is 0°, the average self-confidence degree of the subjects is more than 3; when the spatial angle is more than 30°, the average self-confidence degree is less than 3. However, female respondents were much more confident at a spatial angle of 0° than male respondents, and there was a slight upward shift in confidence at a spatial angle of 80°. Overall, with the increase of the spatial angle, the degree of confidence of the subjects in the selection decreased, and the degree of confidence in the subjects' selection was the same when the spatial angle was 70° and 80°. From the perspective of spatial, the influence of the subject's choice confidence level is analyzed when the spatial perspective changes, the female's choice confidence level will decrease.

4) Spatial direction information transfer efficiency and capability analysis

The simulated data of experiment 1 of the hangtag-type safety evacuation signs show that the overall response time of the subjects is positively correlated with the spatial angle. The expected selection conformity and the degree of selection confidence are negatively correlated with the spatial angle, but the degree of confidence increases slightly when the spatial angle is 80°. This is because when the spatial angle is 80°, the subjects observed that the safety evacuation sign is the same as when people are located on the side of the safety evacuation sign in real life. Due to inertial thinking[52], the subjects will mistakenly believe that the safety evacuation signs at this time are the same as those in their own memory and respond based on this. The overall choice response time is shortened, the confidence level increases, but the correct rate is lower.

When the space angle of the safety evacuation sign changes, the area where the subject observes the safety evacuation sign will also change, which affects the respondent's response time and the accuracy of the selection. Therefore, using the visible area can more intuitively explain the influence of the change of the spatial angle of the safety evacuation sign on its spatial direction information transmission efficiency and capability. Then $S_{d\alpha}$ is calculated as follows.

$$S_{d0} = ab \qquad (1)$$

$$K_d = sin（180°- \arctan（L/H）\pm \alpha） \qquad (2)$$

$$S_{d\alpha} = S_{d0} \times K_d = ab sin（180°- \arctan（L/H）\pm \alpha） \qquad (3)$$

Note: When the safety evacuation sign is away from the evacuees, take +, otherwise take -.

In the formula, $S_{d\alpha}$ is the projected area of the hangtag-type safety evacuation sign that the tested person can see from the front, the visible area. $\alpha$ is the spatial angle. $K_d$ is the visible area ratio, equal to the ratio of the visible area $S_{d\alpha}$ when the spatial angle is $\alpha$ to the maximum value of the visible area $S_{d0}$ when the spatial angle is 0°. The embedded safety evacuation sign is a in length and b in width. The vertical distance between the tested person and the embedded safety evacuation sign is H, and the straight-line distance is L.

In order to calculate the visible area of the hangtag-type safety evacuation sign, it is known that the commonly used hangtag-type safety evacuation sign in China is 359mm long and 149mm wide. Combined with the actual passage width of the site and the optimal observation distance, set the width of the safe passage and the linear distance L between the tested person and the safety evacuation sign to be 2000mm. Then the vertical distance H between the tested person and the safety evacuation sign is (0, 1000mm). The calculation results are shown in Tab.1.

Tab.1 Influence of space angle of hangtag-type safety evacuation sign on the visible area

| Space angle | 0° | 30° | 60° | 70° | 80° |
|---|---|---|---|---|---|
| Range of visible area /mm² | 47844-53491 | 29472-53394 | 3304-44638 | 0-38842 | 0-31866 |
| Visible area ratio | 1 | 0.99 | 0.83 | 0.73 | 0.60 |

It can be seen from Tab.1 that when $S_{d0}$ remains unchanged, the visible area of the hangtag-type safety evacuation sign is inversely proportional to the spatial angle. When the spatial angle is 80°, the minimum visible area is 0mm², the maximum is 31866mm², and the visible area ratio is only 0.60, which is much smaller than the visible area of 53491mm² when the spatial angle is 0°. Combined with the data analysis of Experiment 1, it can be seen that the influence degree of the subjects' choice response after seeing the evacuation sign, the degree of agreement of the expected choice, and the degree of confidence in the choice are related to the visible area of the hangtag-type safety evacuation sign. When the visual area is larger, the choice response time is shorter, and the expected choice conformity and confidence are higher.

### 3.2.2 Experiment 2

The simulation results and analysis of experiment 1 show that the information transfer efficiency and capability gradually weaken with the increase of the spatial angle. To verify the validity of the experimental results, take 0° as the control angle, select 70° and 80° with the longest average response time of the subjects to carry out experiment 2 (the embedded safety evacuation sign spatial angle is 0° when the hangtag-type sign type safety evacuation sign spatial angle changes). After the Friedman M test, when the spatial angle is 0°, there is no statistically significant difference between the data in the group. When the spatial angle is 70°, there is a significant difference in the evacuation time between Exit 1 and Exit 4 (p=0.001); when the spatial angle is 80°, there is a significant difference in the evacuation time between Exit 2 and Exit 4 (p=0.013). The specific analysis is as follows.

1) Escape exit

Fig.22 shows the effect of spatial angle on participants' choice of escape exit.

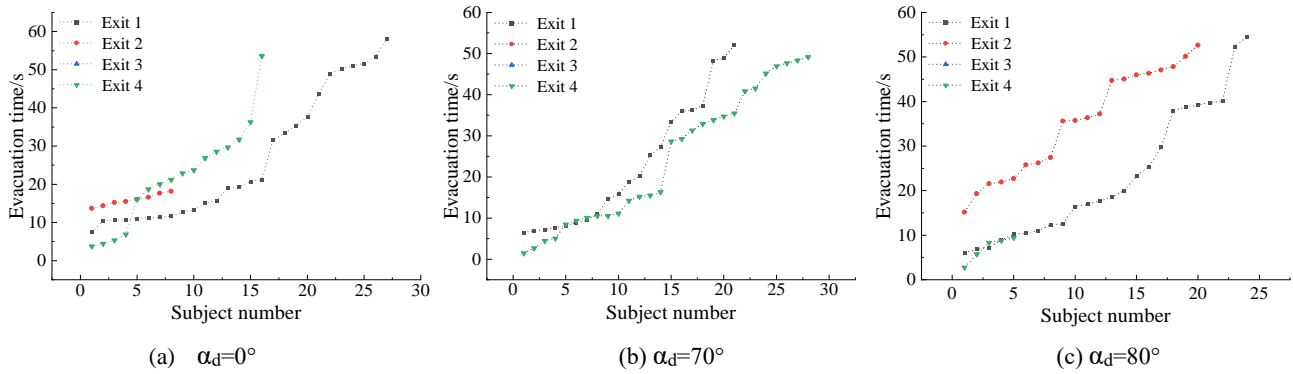

(a) $α_d=0°$  (b) $α_d=70°$  (c) $α_d=80°$

**Fig.22 Comparison of escape exit selection with spatial angle**

As shown in Fig.22(a), when the spatial angle is 0°, the most significant number of participants escaped from Exit 1, and the participant with the longest escape time escaped from Exit 1, and no participants escaped from Exit 3. In Fig.22(b), when the spatial angle is 70°, the only escape exit selected by the participants are Exits 1 and 4. The participants who escaped from Exit 4 were slightly higher than Exit 1 and had the longest escape time. The participants escaped from Exit 1. Fig.22(c) shows that the spatial angle of 80° is consistent with the spatial angle of 0°, the most significant number of participants escaped from Exit 1, and the participants with the longest escape time escaped from Exit 1, and no participants escaped from Exit 3. This is because the participants are more familiar with Exit 1 and are more willing to escape to the exits they are familiar with during the escape process[53].

2) Evacuation time

Tab.2 shows the comparison of the evacuation time of the participants at each safety exit. Among them, Sum is the sum of the escape time used by the escape personnel at each exit. Total is the sum of the number of escaped persons and the sum of escape time from each space angle. Avg refers to the average value of the escape time errors used by escaping personnel at each exit. Mean refers to the average value of escape time at each spatial angle.

**Tab.2 Fire drill evacuation time comparison**

| Space angle | Escape time /s | Exit 1 | Exit 2 | Exit 3 | Exit 4 | Total | Mean |
|---|---|---|---|---|---|---|---|
| 0° | Sum | 716.54 | 127.26 | 0 | 349.95 | 1193.75 | |
| | Num | 27 | 8 | 0 | 16 | 51 | |
| | Avg | 26.54 | 14.14 | 0 | 21.87 | | 23.41 |
| 70° | Sum | 387.28 | 0 | 0 | 620.62 | 1007.90 | |
| | Num | 17 | 0 | 0 | 25 | 42 | |
| | Avg | 22.78 | 0 | 0 | 24.82 | | 24.00 |
| 80° | Sum | 469.48 | 603.68 | 0 | 21.06 | 1094.22 | |
| | Num | 20 | 17 | 0 | 3 | 40 | |
| | Avg | 23.47 | 35.51 | 0 | 7.02 | | 27.36 |

It can be seen from Tab.2 that when the spatial angle is 0°, the average escape time of the participants is 23.41s; when the spatial angle is 70°, the average escape time of the participants is 24.00s; when the spatial angle is 80°, the average escape time of the participants is 27.36s. Most participants chose exits 1 and 4 to escape, and no participants escaped from exit 3. However, when the spatial angle was 80°, 42.5% of the participants escaped from Exit 2. This is because when the spatial angle of the safety evacuation sign is 80°, the participants can hardly see the front of the fire safety sign and thus misunderstand it. As a result, the participants' safety exit selection is inconsistent with the space angle of 0° and 70°. It can be seen that the appropriate spatial angle is helpful to enhance the information transmission efficiency and capability of the safety evacuation sign's spatial direction and improve the emergency evacuation speed of the participants.

### 3.3 Embedded safety evacuation signs
#### 3.3.1 Experiment 1
1) Straight safety channel

Fig.23 shows the effect of spatial angle on the subjects' choice of response.

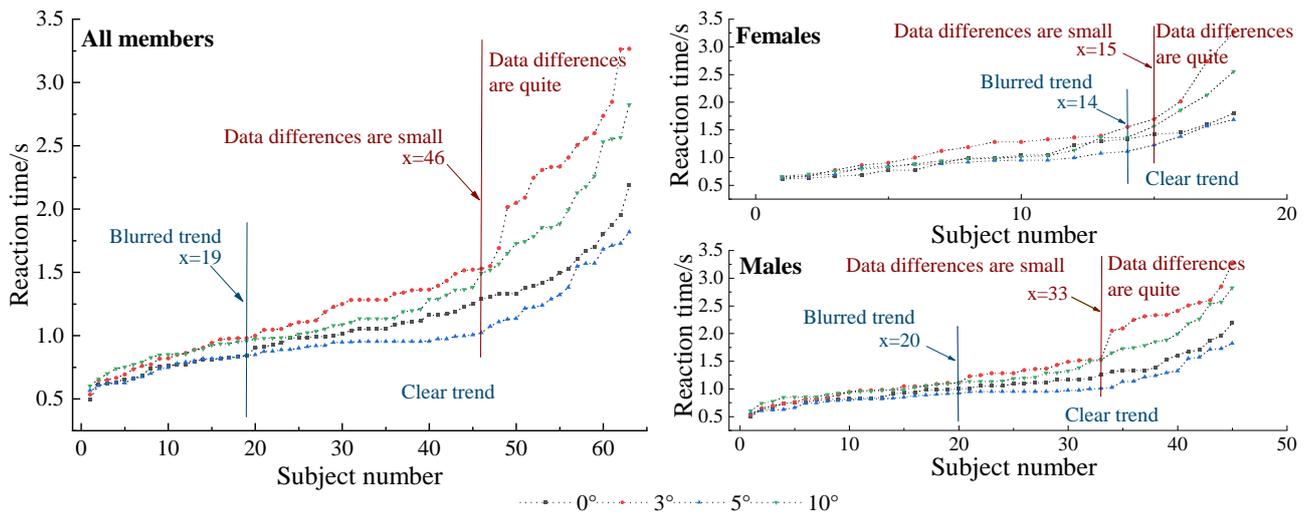

Fig.23 Selecting the response time to change with space angle contrast

The results of each spatial angle pairing in this part of the experimental data are statistically significantly different (p=0), among which, when 0°vs5°, p=0.023, and when 3°vs10°, p=0.035. As shown in Fig.23, when all members are analyzed, when the spatial angle is 5°, the subjects select the shortest response time. When the spatial angle is 3°, the subjects respond the longest, and when the number of subjects exceeds 19, this trend is more straightforward. Among them, 17 people's choice response data are quite different from others. When analyzed by gender, the trend is consistent with the overall analysis. When the number of male subjects is more than 20 and the number of female subjects is more than 14, the trend is more obvious. 12 of the male subjects and 3 of the female subjects reacted significantly differently from the others. When the spatial angle is 3°, the safety evacuation sign is different from what the tested person sees in daily life, but it is not possible to obtain the spatial direction information conveyed by it. When the space angle is 10°, the subjects observed that the safety evacuation signs are quite different from real life, affecting the subjects' judgment. Judging from the spatial perspective on the influence of the subjects when they selected to respond, there is little difference between males and females.

Fig.24 shows the effect of spatial angle on the subject's choice of direction.

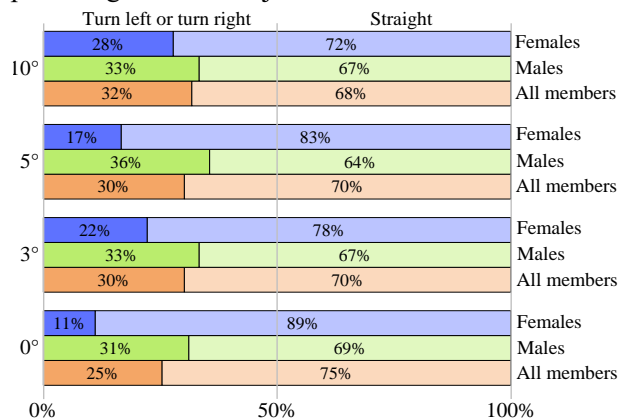

Fig.24 Direction selection and space angle quantitative relationship

Only the straight (intended direction) direction is available in the embedded safe evacuation sign evacuation scenario. As shown in Fig.24, when all members were analyzed, more than 70.75% chose the expected direction. Among them, when the space angle is 10°, the proportion of people who choose the expected direction is the lowest, only 68%, and when the space angle is 0°, the proportion of people who select the expected direction is the highest, which is 75%. When analyzed by gender, more than 66.75% of male respondents chose the expected direction. Among them, when the space angle is 5°, the proportion of people who choose the expected direction is the lowest, only 64%, and when the space angle is 0°, the proportion of people who select the expected direction is the highest, 69%. More than 80.5% of the female respondents chose the expected direction. Among them, when the space angle is 10°, the proportion of people who choose the expected direction is the lowest, only 72%, and when the space angle is 0°, the proportion of people who select the expected direction is the highest, at 89%. From the effect of spatial angle on the tester's direction choice, it can be seen that the proportion of women choosing the expected direction is higher than that of men.

Fig.25 shows the effect of spatial angle on the subjects' choice confidence.

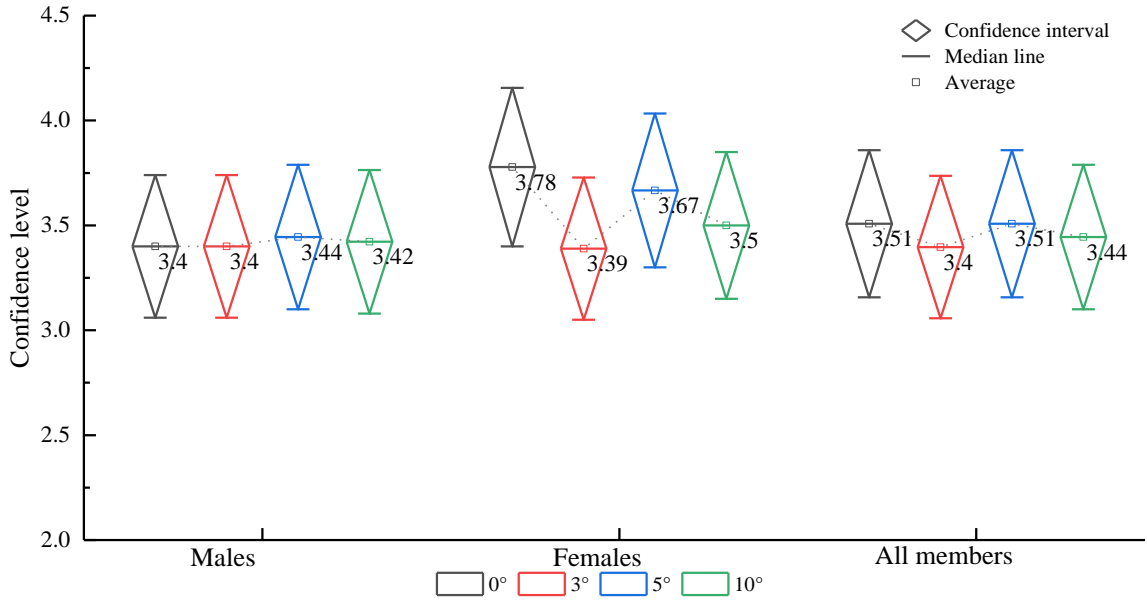

**Fig.25 Degree of confidence and space angle quantitative relationship**

As shown in Fig.25, there is little difference in the degree of self-confidence of the subjects, with a mean value of more than 3, indicating that the subjects are confident in their own choices under the conditions of 4 different spatial angles. Analyzing all members, when the spatial angle is 5°, the subjects are slightly more confident in their choice than in other situations. When analyzed by gender, male respondents were slightly more confident in their choice at a spatial angle of 5° than in other cases; female respondents were slightly more confident in their choice at a spatial angle of 0° than in other cases at a spatial angle of 5°. Female respondents were overall more confident in their choice than male respondents. The effect of spatial angle on the testers' confidence level of choice shows that the women's confidence level decreases when the spatial angle changes.

2) Spatial direction information transfer efficiency and capability analysis

The experimental simulation data of experiment 1 of the embedded safety evacuation sign shows that when the space angle is 5°, the subject selects the shortest response time. When the space angle is 3°, select the longest response time. On the whole, more than 70% of the people chose the expected direction, and the average self-confidence degree of the subjects was more than 3. When the spatial angle was 5°, the subjects were slightly more confident in their choice than in other cases.

Then the visible area $S_{q\alpha}$ of the embedded safety evacuation sign is calculated as follows:

$$S_{q0} = ab \tag{4}$$

$$K_q = \frac{\sqrt{(H^2+L^2)} \times sin(180° - arctan(H/L) - \alpha)}{\sqrt{(H-asin\alpha)^2 + (L+acos\alpha)^2}} \tag{5}$$

$$S_{q\alpha} = S_{q0} \times K_q = ab \frac{\sqrt{(H^2+L^2)} \times sin(180° - arctan(H/L) - \alpha)}{\sqrt{(H-asin\alpha)^2 + (L+acos\alpha)^2}} \tag{6}$$

In the formula, $S_{q\alpha}$ is the projected area of the embedded safety evacuation sign that the tested person can see from the front, the visible area. α is the spatial angle. $K_q$ is the visible area ratio, equal to the ratio of the visible area $S_{q\alpha}$ when the spatial angle is α to the maximum value of the visible area $S_{q0}$ when the spatial angle is 0°. The embedded safety evacuation sign is a in length and b in width. The vertical distance between the tested person and the embedded safety evacuation sign is H, and the straight-line distance is L.

In order to calculate the visible area of the embedded safety evacuation sign, it is known that the embedded safety evacuation sign is 359mm long and 149mm wide. Combined with the actual aisle width of the site and the optimal observation distance, the width of the safe passage and the straight line between the measured person and the safety evacuation sign are set. If the distance L is 2000mm, the vertical distance H between the tested person and the safety evacuation sign is (0, 2000mm). The calculation results are shown in Tab. 3.

Tab.3 Influence of space Angle of embedded safety evacuation sign on the visible area

| Space angle | 0° | 3° | 5° | 10° |
|---|---|---|---|---|
| Range of visible area /mm$^2$ | 0-1927 | 101-2028 | 168-2094 | 335-2247 |
| Visible area ratio | 1 | 1.05 | 1.09 | 1.17 |

It can be seen from Tab.3 that when $S_{q0}$ remains unchanged, the visible area of the embedded safety evacuation sign is proportional to the spatial angle. When the spatial angle is 0°, the maximum visible area is 1927mm$^2$, which only accounts for 3.6% of the area of the safety evacuation sign. The visual appeal is weak, especially when people walk close to the wall; it is challenging to observe the conveyed spatial direction. When the space angle is 5°, the maximum visible area increases to 2094mm$^2$, and when the space angle is 10°, the maximum visible area is 2247mm$^2$. Combined with the analysis of the data in experiment 1, it can be seen that the response time, the expected choice conformity, and the degree of confidence in the choice made by the subjects after seeing the evacuation signs are related to the visible area of the embedded safety evacuation signs the more significant the visible area is, the shorter the response time and the higher the expected choice conformity and the degree of confidence in the choice. However, when the spatial angle is 10°, the subjects observed that the embedded safety evacuation signs are quite different from real life, affecting the subjects' judgment. Therefore, when the spatial angle is 5°, the embedded safety evacuation signs are in the spatial direction information transmission ability is the strongest, which shortens the time when the tested person selecting to respond, thereby reducing the emergency evacuation time in the event of a fire and improving the evacuation efficiency.

### 3.3.2 Experiment 2

The experiment 1 simulation results and analysis show that when the spatial angle is 5°, the spatial direction information transfer efficiency and capability are the best, and the daily embedded safety evacuation sign spatial angle is 0°. Therefore, the spatial angle of 0° and 5° is selected for experiment 2 (when the embedded safety evacuation sign spatial angle changes, the hangtag-type safety evacuation sign spatial angle is 0°) to verify the experiment 1 results. After the Friedman M test, when the spatial angle was 0° and 5°, there was no statistically significant difference between the data within the group.

1) Escape exit

Fig.26 shows the effect of spatial angle on participants' choice of escape exit.

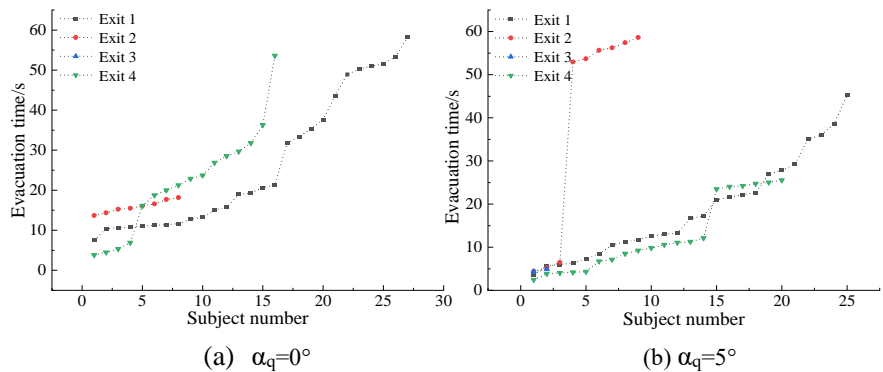

(a) $α_q=0°$　　(b) $α_q=5°$

Fig.26 Comparison of escape exit selection with spatial angle

As shown in Fig.26(a), when the spatial angle is 0°, the largest number of participants escaped from Exit 1. The participant with the longest escape time escaped from Exit 1, and no participants escaped from Exit 3. In Fig.26(b), the most significant number of participants escaped from Exit 1, the least number of participants from Exit 3, only 2 people, and the participant with the longest escape time escaped from Exit 2. Consistent with the second experiment of the hangtag-type safety evacuation sign, the evacuees are more willing to escape from the more familiar Exit 1. However, when the spatial angle is 5°, the proportion of participants choosing non-familiar exits increases, and participants are escaping from Exit 3, which no one decides when the spatial angle is 0°. It shows that when the spatial angle is 5°, the visible area of the embedded safety evacuation sign is increased, the information transmission efficiency and capability of the spatial direction are enhanced, and the visual appeal to the participant's increases.

2) Evacuation time

Tab.4 shows the comparison of the evacuation time of the participants at each safety exit. Consistent with Experiment 1, Sum is the sum of the escape times used by escapees from all exits. Total refers to the sum of the number of people who escaped from each space angle and the sum of the escape time. Avg refers to the average value of the escape time errors used by escaping personnel at each exit. Mean refers to the average value of escape time at each spatial angle.

Tab.4 Fire drill evacuation time comparison

| Space angle | Escape time /s | Exit 1 | Exit 2 | Exit 3 | Exit 4 | Total | Mean |
| --- | --- | --- | --- | --- | --- | --- | --- |
| 0° | Sum | 716.54 | 127.26 | 0 | 349.95 | 1193.75 | |
| | Num | 27 | 8 | | 16 | 51 | |
| | Avg | 26.54 | 14.14 | 0 | 21.87 | | 23.41 |
| 5° | Sum | 469.97 | 350.80 | 9.28 | 252.32 | 1082.37 | |
| | Num | 25 | 9 | 2 | 20 | 56 | |
| | Avg | 18.80 | 38.98 | 4.64 | 12.616 | | 19.33 |

It can be seen from Tab.4 that when the spatial angle is 0°, the average escape time of the participants is 23.41s, and when the spatial angle is 5°, the average escape time of the participants is 19.33s. Among the exits with the most escape choices for the 2 participants, No. 1 and No. 4, the average escape time when the spatial angle is 5° is lower than when the spatial angle is 0°. In the 2 exercises, the participants who escaped from Exit 1 accounted for more than half of the total number of participants, and when the spatial angle was 5°, each safety exit had participants escape. It shows that when the spatial angle is 5°, the participants can evacuate faster under the guidance of embedded safety evacuation signs, and at the same time, use safety passages more rationally.

## 4  Safety evacuation sign design

It can be seen from the above analysis that whether it is a hangtag-type or an embedded safety evacuation sign, the information transmission efficiency and capability of the spatial direction are affected by the spatial angle. At present, the typical safety evacuation signs are all flat structures. When people's line of sight is parallel to them, the spatial direction information they convey cannot be observed, which will prolong the response time of evacuees and reduce the evacuation efficiency. Therefore, increasing the visual appeal of safety evacuation signs to make them more recognizable has a positive effect on improving evacuation efficiency. Based on this, this paper combines the experimental results to design space three-dimensional structure safety evacuation signs so that people can observe the spatial direction information conveyed from different angles and increase the efficiency and capability of its spatial direction information transmission to improve the efficiency of an emergency evacuation.

1) Design of hangtag-type safety evacuation sign in the three-dimensional spatial structure

The data of hangtag-type safety evacuation sign experiment 1 and experiment 2 show that the spatial direction information transmission efficiency and capability are negatively correlated with the spatial angle. The smaller the visible area of the safety evacuation sign, the worse the spatial direction information transmission efficiency and capability. Moreover, the common hangtag-type safety evacuation signs only have two sides. When people are located in different positions, the viewing angles are different, which will cause significant interference to the acquisition of spatial direction information. This paper proposes a polyhedral hangtag-type safety evacuation sign (as shown in Fig.27). The advantages of this kind of structural space direction information transmission mainly lie

in:

a. It can display the spatial direction information on eight sides so that people can more effectively obtain the spatial direction information it conveys from different angles.

b. Safety evacuation signs can be observed in the upper and lower parts, which can be adapted to people of different heights.

c. Increase the efficiency of spatial direction information transmission and improve the efficiency of an emergency evacuation.

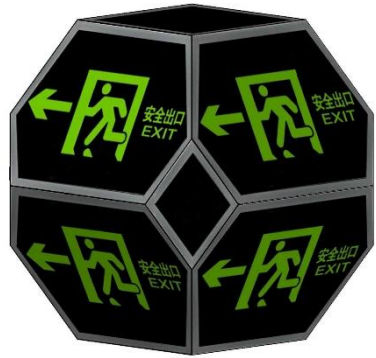

**Fig.27  Three-dimensional spatial structure hangtag-type safety evacuation sign three-dimensional view**

2) Design of embedded safety evacuation sign in the three-dimensional spatial structure

The data analysis results of embedded safety evacuation sign experiment 1 show that: when the spatial angle is 5°, the subjects select the shortest response time, the expected choice coincidence degree is 84.48%, and the choice confidence degree is the highest, which is 3.62; in experiment 2, the spatial angle is When the spatial angle is 5°, the average escape time of the participants is 19.328s, which is lower than the average escape time of 23.4068s when the spatial angle is 0°; that is, when the spatial angle is 5°, the spatial direction information transmission efficiency and capability of the embedded safety evacuation sign is the best. Based on this, this paper designs a spatial three-dimensional structure embedded safety evacuation sign with the angle between the left and right slopes and the wall of 5° (as shown in Fig.28). As shown below, this structured space direction information transmission has four main advantages.

a. The left and right slopes of the quadrangular prism shell and the mounting plate form an included angle of 5°.

b. The four sides display the spatial direction information.

c. The choice of the spatial angle of 5° makes the three-dimensional spatial structure embedded. The safety evacuation sign will not take up too much space while highlighted.

d. Increase the visible area of the embedded safety evacuation sign, reduce the response time of evacuees, to determine the direction of the safety exit and the distance to the nearest safe exit in time so that people can quickly get from the nearest safe exit escape.

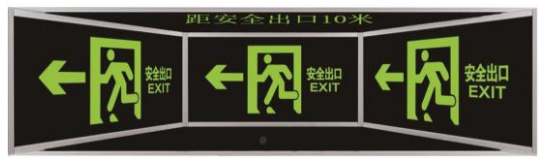

**Fig.28  Three-dimensional spatial structure embedded safety evacuation sign three-dimensional view**

The existing fire safety evacuation is a plain design with a relatively single function, which can only display the direction information on the sign surface. So, the direction information can be projected to the ground or roof (as shown in Fig.29) to increase the visible area of the safety evacuation signs, reduce the response time of evacuees, and improve the efficiency of an emergency evacuation.

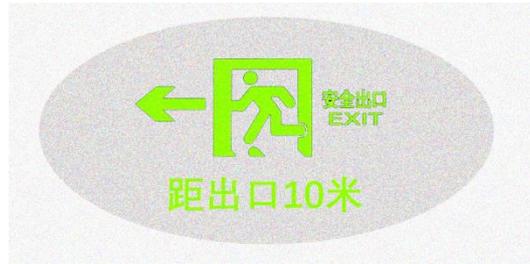

**Fig.29 Projection diagram of evacuation sign in the three-dimensional structure**

# 5 Discussion

Effective safety evacuation signs can correctly transmit spatial direction information, and they have different attractiveness to evacuees when they are in different spatial positions. The more the safety evacuation signs attract the attention of evacuees, the more timely and effective people can obtain the space direction information indicated by them, choose the correct evacuation route, and reduce the evacuation time. Most previous studies on safe evacuation signs only use fire drills[54][55] or simulation experiments[39][56][57]. Although fire drills can restore evacuation scenarios to a greater extent, data collection is more difficult. Although the simulation test can use smart devices to obtain more accurate data, it is far from the actual evacuation scene. Therefore, this paper adopts a combination of simulation experiments and fire drills to obtain more accurate results.

The research on the information transmission efficiency of the safety evacuation signs in the space direction provides a new idea for a better understanding of the influence of safety evacuation signs on the selection of evacuation routes. Based on the research of Kubota et al.[40], this paper adds to research on embedded safety evacuation signs and research on different genders' effects and validates it through fire drills. The experimental results show that the spatial angle of the safety evacuation signs will affect the response time of evacuees, the confidence level and accuracy of the selected route, and have different effects on evacuees of different genders.

In Experiment 1, the distance between the subjects and the observed safety evacuation signs and the line of sight is fixed, which is the first step to understanding better the interaction between the embedded safety evacuation sign spatial structure and the evacuees. The spatial relationship between evacuees and safety evacuation signs is constantly changing in the fire evacuation process. Although it has been verified by experiment 2, there are no statistics on the sight placement of the exercisers. Future research should consider the influence of the spatial angle of embedded safety evacuation signs on their spatial direction information transfer efficiency and capability under the dynamic changes of evacuees. The research subjects are all young students, and the research is divided by gender. However, age, occupation, personal psychology, and familiarity with the environment will affect how people find their direction and position themselves in the environment[58][59]. These individual factors should be more fully considered in subsequent studies to improve the generalizability of the results. Constrained by the exercise site, the sample size of Experiment 2 is small, and accidental errors in the data are unavoidable, which should be improved in the later research. At the same time, when vision is affected, it is more difficult for evacuees to accept the spatial direction information transmitted by embedded safety evacuation signs. Therefore, factors affecting vision such as smoke and dust need to be considered. In addition, the visual selection response time experiment shows that the response time of different genders is different, and experiment 1 also indicates that the spatial angle change of the safety evacuation sign has other effects on the subjects of different genders. Follow-up research can also focus on this point, do a more in-depth study. In order to further study the relationship between the spatial angle of the safety evacuation sign and the information transfer efficiency and capability of the spatial direction, this paper proposes a formula for calculating the visible area of the safety evacuation sign. Although Kubota et al.[40] also obtained a similar conclusion by calculating the visible area, "the larger the visible area, the shorter the selection response time, and the higher the expected selection consistency and confidence". However, in this paper, the visual area is used for analysis rather than the visual area calculated for a fixed position, and a more intuitive, detailed, and accurate conclusion can be obtained. According to the experimental results, this paper also designs the safety evacuation signs of the hangtag-type and the embedded space three-dimensional structure, which provide ideas for the subsequent application and promotion of the experimental conclusions.

## 6 Conclusion

This paper analyzes the degree of influence of 2 types of common safety evacuation signs, hangtag-type and embedded, on subjects' choice response time, expected choice conformity, and choice confidence level by designing a spatial directional information transfer efficiency and capability simulation experiment and conducts a more in-depth analysis from gender. In experiment 1 of the hangtag-type safety evacuation sign, the response time of the subjects was positively correlated with the spatial angle as a whole, the expected choice conformity and choice confidence were negatively correlated with the spatial angle, that is, the spatial direction information transmission efficiency and capability were negatively correlated with the spatial angle, Experiment 2 further verified this result. In the embedded safety evacuation sign experiment 1, when the spatial angle is 5°, the spatial direction information transmission efficiency and capability is the strongest, and the subjects selected respond in the shortest time. In experiment 2, when the spatial angle is 5°, the participants escape on average time is shorter, the proportion of participants selecting unfamiliar exits increases, and the choices for safe exits are more diverse. When analyzed by gender, it was found that there was little difference in the choice of response by gender; males were more likely than females to be influenced by changes in spatial angle when making directional choices, and females were more likely to be influenced by spatial angle when choosing their level of confidence. The difference test of the data of experiment 2 shows that the change of the spatial angle of the hangtag-type safety evacuation sign has a more obvious impact on the evacuees. The spatial three-dimensional structure safety evacuation sign designed in this paper can display the spatial direction information in multiple directions and has a projection function, which can increase the visible area of the embedded safety evacuation sign, improve the spatial direction information transmission efficiency and capability, and reduce the response time of evacuees. It provides a new direction for improving the functional structure of safe evacuation signs. This conclusion can be used to improve the design and installation specifications of fire safety signs and provide new ideas for more effective use of safety evacuation signs. Future research can further explore the interaction between people and safety evacuation signs and the influence of different individual characteristics on the interaction between people and signs.

## Declaration of Competing Interest

The authors declare that they have no known competing financial interests or personal relationships that could have appeared to influence the work reported in this paper.

## Acknowledgments

This work was supported in part by the Science and Technology Plan Project of Guizhou Province, China [2020]4Y055, and in part by the Science and Technology Plan Project of Guizhou Province, China (grant number [2022] 250).